\documentclass[twocolumn,trackchanges]{aastex63}
\usepackage{amsmath}
\usepackage{graphicx}
\usepackage{amsfonts}
\usepackage{natbib}
\usepackage{color}
\usepackage{epstopdf}
\usepackage{graphicx}
\usepackage{multirow}
\usepackage{longtable}
\usepackage{rotating}

\shorttitle{$\gamma$-ray emission from 1LHAASO J0249+6022}
\shortauthors{Gong et al.}

\begin{document}

\title{Detection of the Extended $\gamma$-ray Emission around TeV source 1LHAASO J0249+6022 with Fermi-LAT}

\author{Yunlu Gong}
\affil{Department of Astronomy, School of Physics and Astronomy, Key Laboratory of Astroparticle Physics of Yunnan Province, Yunnan University, Kunming 650091, People's Republic of China; lizhang@ynu.edu.cn, fangjun@ynu.edu.cn}

\author{Liancheng Zhou}
\affil{Department of Astronomy, School of Physics and Astronomy, Key Laboratory of Astroparticle Physics of Yunnan Province, Yunnan University, Kunming 650091, People's Republic of China; lizhang@ynu.edu.cn, fangjun@ynu.edu.cn}

\author{Qi Xia}
\affil{Department of Astronomy, School of Physics and Astronomy, Key Laboratory of Astroparticle Physics of Yunnan Province, Yunnan University, Kunming 650091, People's Republic of China; lizhang@ynu.edu.cn, fangjun@ynu.edu.cn}

\author{Shan Chang}
\affil{Department of Astronomy, School of Physics and Astronomy, Key Laboratory of Astroparticle Physics of Yunnan Province, Yunnan University, Kunming 650091, People's Republic of China; lizhang@ynu.edu.cn, fangjun@ynu.edu.cn}

\author{Jun Fang}
\affil{Department of Astronomy, School of Physics and Astronomy, Key Laboratory of Astroparticle Physics of Yunnan Province, Yunnan University, Kunming 650091, People's Republic of China; lizhang@ynu.edu.cn, fangjun@ynu.edu.cn}

\author{Li Zhang}
\affil{Department of Astronomy, School of Physics and Astronomy, Key Laboratory of Astroparticle Physics of Yunnan Province, Yunnan University, Kunming 650091, People's Republic of China; lizhang@ynu.edu.cn, fangjun@ynu.edu.cn}



\begin{abstract}
1LHAASO J0249+6022 is an extended very-high-energy gamma-ray source discovered by the Large High-Altitude Air Shower Observatory. Based on nearly 16.1 years of data from the Fermi Large Area Telescope, we report the probable gamma-ray emission from 1LHAASO J0249+6022 in the 0.03-1 TeV energy range. The results show that its gamma-ray spectrum can be well fitted by a single power law with an index of 1.54 $\pm$ 0.17, and integral photon flux is (4.28 $\pm$ 1.03) $\times$ 10$^{-11}$ photons cm$^{-2}$ s$^{-1}$. We also considered theoretically whether the non-thermal emission could originate from a pulsar wind nebula (PWN) scenario. Assuming that the particles injected into the nebula have a power-law distribution, the resulting spectrum from the inverse Compton scattering is consistent with the detected GeV and TeV gamma-ray fluxes. Our study shows that the PWN scenario is reasonable for 1LHAASO J0249+6022.

\end{abstract}


\section{Introduction}
\label{sec:intro}
Efficient particle accelerators responsible for Galactic cosmic rays are abundant in our Galaxy and are capable of producing high-energy $\gamma$-ray through various emission mechanisms, such as inverse Compton scattering on ambient photon fields and/or the bremsstrahlung process of high-energy electrons (leptonic model) and the decay of neutral pion mesons produced by hadronic collisions with ambient material (hadronic model). This means that $\gamma$-ray observations play a crucial role in understanding non-thermal astrophysical processes. A population of $\gamma$-ray sources has been discovered by several deep sky observations, such as the Fermi Large Area Telescope (Fermi-LAT), the High Energy Stereoscopic System (H. E. S. S.), the High Altitude Water Cherenkov (HAWC), and the Large High-Altitude Air Shower Observatory (LHAASO), some of which have been identified as pulsar wind nebulae (PWNe), supernova
remnants (SNRs), young star clusters, TeV halos, and so on \citep{2018A&A...612A...1H,2020ApJS..247...33A,2020ApJ...905...76A,2024ApJS..271...25C}. However, a large proportion of the source classes have not yet been identified and their multi-band counterparts need to be investigated in more detail. Multi-band observations are essential for understanding the nature of the source and for studying how and where cosmic rays are accelerated \citep{2020ApJ...904..123E,2024ApJ...965...28G}. Currently, a series of multi-band searches have been carried out, e.g. MGRO J1908+06 \citep{2021ApJ...913L..33L}, G272.2-3.2 \citep{2021ApJ...918...24X}, N 157B \citep{2023MNRAS.526..193G}, G32.64+0.53 \citep{2024ApJ...972...84X}, CTA 1 \citep{2024MNRAS.529.3593Z}, and 1LHAASO J1945+2424 \citep{2024MNRAS.527.8006A}.

In the first catalogue of very-high-energy and ultra-high-energy $\gamma$-ray sources, 32 new TeV sources (e.g. 1LHAASO J0249+6022) have been announced based on the association criterion \citep{2024ApJS..271...25C}. 1LHAASO J0249+6022 is an extended source detected simultaneously by (39\% containment radius $r_{39}\approx0.71^{\circ}$) WCDA and ($r_{39}\approx0.38^{\circ}$) KM2A detectors, with a positional offset of $\sim0.45^{\circ}$ \citep{2024ApJS..271...25C}. Next, \cite{2024arXiv241004425C} observed excess $\gamma$-ray induced showers using the LHAASO-WCDA data of live 796 days and LHAASO-KM2A data of live 1216 days. In the case of a two-dimensional Gaussian model, the best-fit position derived from WCDA data is $\mathrm{R.A.}=42.06^\circ\pm0.12^\circ $ and $\mathrm{Dec.}=60.24^{\circ}\pm0.13^{\circ}$ with an extension of $0.69^{\circ}\pm0.15^{\circ}$, while for KM2A data, it is $\mathrm{R.A.}=42.29^\circ\pm0.13^\circ $ and $\mathrm{Dec.}=60.38^{\circ}\pm0.07^{\circ}$ with an extension of $0.37^{\circ}\pm0.07^{\circ}$. For extended TeV $\gamma$-ray emission, the inverse Compton process of highly relativistic electrons and positrons was considered to be well explained by \cite{2024arXiv241004425C}.

1LHAASO J0249+6022 is in positional coincidence with the energetic pulsar PSR J0248+6021. The pulsar was first discovered during a survey of the northern Galactic plane with the Nancay radio telescope, and was later detected in the GeV band by the Fermi-LAT \citep{1997AAS...19111110F,2010ApJS..187..460A}. The pulsar has a rotation period of 0.217 s and a first period derivative of $5.51\times10^{-14}\text{s s}^{-1}$ \citep{2011ApJ...733...82M}. These parameters lead to a characteristic age of $\sim$63 kyr and a spin-down luminosity $L(t) = 2.04\times10^{35} \mathrm{erg} \ \mathrm{s}^{-1}$. No X-ray emission has been detected around the pulsar, but the upper limit of the unabsorbed X-ray flux is given as $ 9.0\times10^{-13} \mathrm{erg} \ \mathrm{cm}^{-2} \mathrm{s}^{-1}$ in 0.3-10.0 keV using the Swift-XRT data \citep{2011ApJ...733...82M}. The distance of the open cluster IC 1848, located in the giant HII region W5, has been estimated to be 2 kpc based on the main sequence fit of the star cluster and a Galactic kinematic model, and PSR J0248+6021 is almost certainly in W5 \citep{2005A&A...440..403K,2011A&A...525A..94T}. In a study of correlations between pulsar parameters, \cite{2020A&A...644A..73W} found that the nulling fractions of statistically larger pulsars is more closely related to the pulsar period than to the characteristic age or energy loss rate, and that the nulling fraction of PSR J0248+6021 was estimated to be between 0.0\% and 34.7\%. According to \cite{2024ApJ...968..117Z} survey of the SNR catalogue, there is no corresponding SNR in the region of 1LHAASO J0249+6022, which they then considered as a possible association with the TeV halo.

Given the unconfirmed nature of 1LHAASO J0249+6022, we first searched for GeV-TeV emission using Fermi-LAT PASS 8 data, and then explored the possible origins of its $\gamma$-ray emission in conjunction with theoretical modelling. In Section~\ref{data}, we present the process used to analyse the data and the results obtained, which include spatial and spectral analyses. In Section~\ref{results} we discuss the origin of the 1LHAASO J0249+6022 $\gamma$-ray emission with a time-dependent one-zone model. The summary of this work is presented in Section~\ref{sumdis}.

\begin{figure}
\centering
\includegraphics[scale=0.8]{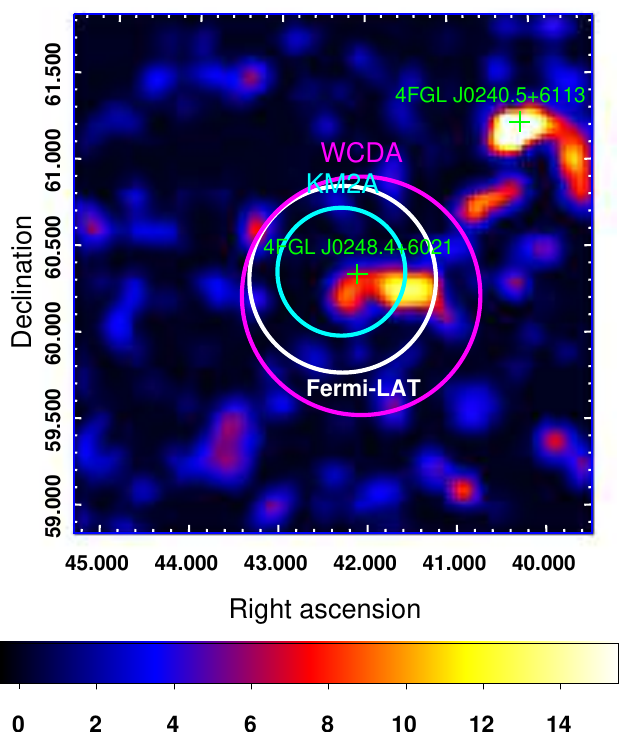}
\caption{TS map for a region of $3_{\cdot}0^{\circ} \times 3_{\cdot}0^{\circ}$ above 30 GeV with Gaussian smoothing of $\sigma = 0.2^{\circ}$. The position of the 4FGL-DR4 sources is indicated by the green cross. The extension of SrcX is marked with the white circle. The cyan and purple circles show the 39\% containment radii of the 2D Gaussian models from KM2A and WCDA sources, respectively \citep{2024arXiv241004425C}.   }
\label{Fig1}
\end{figure}

\begin{figure}
\centering
\includegraphics[scale=0.54]{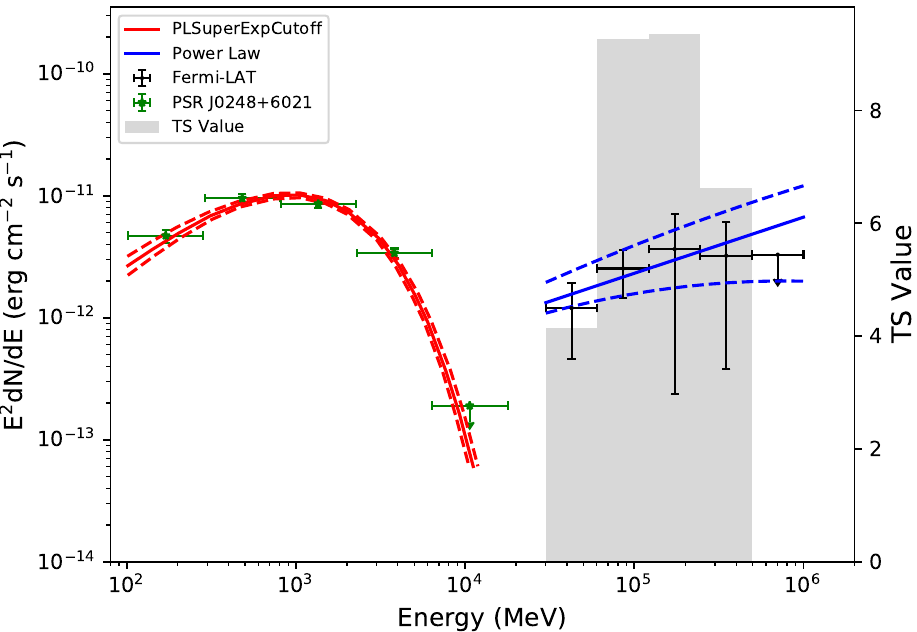}
\caption{The results of the Fermi-LAT spectral data. The black dots indicate the SED of SrcX in the 0.03-1 TeV band, together with the global best-fit spectra (1.54 $\pm$ 0.17) with 1$\sigma$ statistical uncertainty shown as the solid and dashed blue lines, respectively. Energy bins with TS values greater than 4.0 are represented by grey histograms, while energy bin with TS value less than 4.0 give upper limit at the 95\% confidence level. The green pentagram indicates the SED of PSR J0248+6021 in the energy range from 100 MeV to 20 GeV. The global best-fit result and its 1$\sigma$ statistical error are plotted with the red solid and red dashed lines, respectively.
}
\label{Fig2}
\end{figure}


\section{Fermi-LAT Data Analysis}
\label{data}
\subsection{Data Reduction}
Fermi-LAT, a pair-conversion $\gamma$-ray telescope, is sensitive to photon energies from 20 MeV to $\ge$ 1 TeV and has continuously monitored the entire sky every 3 hours \citep{2009ApJ...697.1071A}. To understand the nature of the diffuse $\gamma$-ray emission 1LHAASO J0249+6022 from a multi-wavelength perspective, the Pass 8 data with the `source' event class ({\tt evclass = 128 \& evtype = 3}) are selected from 2008 August 4 (MET 239557418) to 2024 August 31 (MET 746775603). Gamma-ray pulsars are characterized by a soft spectrum, with the flux steeply falling above few GeV, and $\gamma$-ray pulsations from pulsars have usually not been seen above 10 GeV \citep{2020A&A...643L..14M,2021ApJ...912..117X,2023A&A...673A..75A}. Therefore, the energy of the event was cut between 30 GeV and 1 TeV to avoid contamination of PSR J0248+6021, allowing a better test of whether the $\gamma$-rays emission could have come from the PWN scenario \citep{2021Natur.594...33C}. In the Fermi-LAT data analysis of PWN, there are cases where data with an energy range above 30 GeV are used: LHAASO J1908+0621 \citep{2021Natur.594...33C}, N 157B \citep{2023MNRAS.526..193G}, and CTA 1 \citep{2024MNRAS.529.3593Z}. Furthermore, the spectrum of PSR J0248+6021 between the 100 MeV and 20 GeV energy ranges can be described by a power law with subexponential cutoff (PLSuperExpCutoff) model, as demonstrated in Fig.~\ref{Fig2} (green pentagram). We selected events with good time intervals on the basis of standard data quality selection criteria ({\tt DATA\_QUAL>0)\&\&(LAT\_CONFIG==1}). A maximum zenith angle of $90^{\circ}$ is adopted to suppress the contribution from the Earth Limb. For the binned maximum likelihood analysis, we considered a region of interest (ROI) with a radius of $10^{\circ}$ centred at 1LHAASO J0249+6022, and the instrumental response function {\tt P8R3\_SOURCE\_V3} is used. We used the user-contributed script {\tt make4FGLxml.py} to generate the background model, which includes the Galactic diffuse emission ({\tt gll\_iem\_v07.fits}), isotropic diffuse emission ({\tt iso\_P8R3\_SOURCE\_V3\_v1.txt}), and all sources listed in the incremental version of the fourth Fermi-LAT catalog \citep[4FGL-DR4;][]{2022ApJS..260...53A,2023arXiv230712546B}. During the likelihood analysis, the normalizations and the spectral parameters of all sources within $3^{\circ}$ from the center of ROI are set free, together with the normalizations of Galactic and isotropic emission.

\subsection{Spatial Analysis}
The 4FGL-DR4 catalogue lists two $\gamma$-ray sources in the direction of 1LHAASO J0249+6022, with 4FGL J0240.5+6113 and 4FGL J0248.4+6021 corresponding to the high-mass X-ray binary system LS I+61$^{\circ}$303 and PSR J0248+6021, respectively. The $3_{\cdot}0^{\circ} \times 3_{\cdot}0^{\circ}$ Test Statistic (TS) map in the 0.03-1 TeV energy band was first generated with the {\tt gttsmap} command by subtracting the emission from the diffuse backgrounds and all 4FGL-DR4 sources (except 4FGL J0248.4+6021) in the best-fit model, as shown in Figure~\ref{Fig1}. The TS map shows a $\gamma$-ray excess around PSR J0248+6021, which is subsequently labelled SrcX. We then add SrcX to the model file as a point source with a power-law spectrum, and optimise its localisation using the {\tt gtfindsrc} command. The best-fit position and 68\% containment radius for SrcX were determined to be $\mathrm{R.A.}=42.27^\circ$, $\mathrm{Dec.}= 60.34^{\circ}$, and $0.04^{\circ}$, respectively. SrcX has a TS value of 6.4 as a point source, while LHAASO observes significant extended emission in this region. Therefore, a single uniform disk, a two-dimensional Gaussian, and a uniform elliptical disc template are used to test whether SrcX $\gamma$-ray emission extends. For elliptical template, we produced several uniform elliptical discs centred at the position of the best-fit with various semimajor axes from $0.4^\circ$ to $0.6^\circ$ in steps of $0.01^\circ$ and a fixed semiminor axis of $0.23^\circ$. The best fitting parameters for the different templates are listed in Table~\ref{tab1}.

We tested whether ScrX is spatially extended or not as follows. In a study of spatial extensions of 2FGL sources, \cite{2012ApJ...756....5L} explicitly stated that the threshold required to regard a LAT source as extended is $\mathrm{TS_{ext}} > 16$, which has been used in several case studies \citep{2021ApJ...911...49X,2024MNRAS.527.8006A,2024MNRAS.529.3593Z}. The expression for $\mathrm{TS_{ext}}$ is $\mathrm{TS_{ext} = 2(ln \mathcal{L}_{ext} - ln \mathcal{L}_{Pt})}$, where $\mathcal{L}_\mathrm{{ext}}$ and $\mathcal{L}_\mathrm{{Pt}}$ are the maximum likelihood values of the extended template and the point-like model, respectively \citep{2012ApJ...756....5L}. The value of $\mathrm{TS_{ext}}$ between the two-dimensional Gaussian model and point source model is calculated to be 26.6, corresponding to 5.1$\sigma$ extension with one additional degree of freedom. Moreover, we also use the Akaike information criterion \citep[AIC;][]{1974ITAC...19..716A}, $\mathrm{AIC}=2k-2\ln(\mathcal{L})$, to compare different models, where $k$ is the number of free parameters in the model. The Gaussian model with the lowest AIC value ($\Delta{\mathrm{AIC}} = 11.8$) is the preferred model. Based on the comparison of $\mathrm{TS_{ext}}$ and AIC values, we found a significant improvement using the extended Gaussian template over the point source assumption. Meanwhile, the extended $\gamma$-ray emission of SrcX is consistent with that of the WCDA and KM2A sources (Figure~\ref{Fig1}), supporting SrcX as the counterpart of 1LHAASO J0249+6022. Subsequently, all subsequent analyses were performed using a 2D Gaussian with $\sigma$ of $0.54^{\circ}$.

\begin{table*}
   \centering
   \caption{Spatial Analysis Results for SrcX between 30 GeV and 1 TeV}
   \label{tab1}
   \renewcommand\arraystretch{1.0}
  \setlength{\tabcolsep}{3.3mm}
   \begin{tabular}{ccccccc} 
      \hline \hline
Spatial Template & Radius ($\sigma$) & Spectral Index & Photon Flux & TS Value & Degrees   \\
 &	   &  & 10$^{-11}$ ph cm$^{-2}$ s$^{-1}$ & & of Freedom  \\\hline
Point source  & ...  & 2.01 $\pm$ 0.46 & 0.29 $\pm$ 0.55 & 6.42 & 4 \\
Uniform disc 	&  $0.53^{\circ}$ & 1.53 $\pm$ 0.22 & 1.72 $\pm$ 0.59 & 17.89 & 5 \\
2D Gaussian 	&  $0.54^{\circ}$ & 1.54 $\pm$ 0.17 & 4.28 $\pm$ 1.03& 33.02 & 5 \\
Ellipse	&  ($0.50^{\circ}$, $0.23^{\circ}$) & 2.03 $\pm$ 0.21 & 1.62 $\pm$ 0.49& 28.71 & 5 \\\hline
\multicolumn{4}{l}{}
   \end{tabular}
\end{table*}

\begin{table}
   \centering
   \caption{The Energy Flux Measurements from SrcX with a Power Law}
   \label{tab2}
   \renewcommand\arraystretch{1.0}
  \setlength{\tabcolsep}{2.0mm}
   \begin{tabular}{cccc} 
      \hline \hline
E & Band & E$^2$dN(E)/dE & TS Value  \\
 GeV&	GeV &  [10$^{-12}$ erg cm$^2$ s$^{-1}$]& \\\hline
 42.60&	30.00--60.49 & 1.20$\pm$ 0.74 & 4.15\\
 85.90&	60.49--121.97 & 2.53 $\pm$ 1.07 & 9.27\\
 173.21& 121.97--245.95 & 3.65 $\pm$ 3.41 & 9.36\\
 349.25&245.95--495.93 & 3.22 $\pm$ 2.84 & 6.63\\
 686.81&495.93--1000.0 & $<$3.43 & 3.61\\\hline
   \end{tabular}
\end{table}

\subsection{Spectral Analysis}
For the spectral energy distribution (SED) analysis of SrcX, we compared the fitting results of log-parabola and power-law functions based on the best-fit spatial model. The results show that the log-parabola function (TS = 32.83) does not significantly improve the fitting results. With the power-law model, the spectral index of SrcX is fitted to be 1.54 $\pm$ 0.17, and integral photon flux is (4.28 $\pm$ 1.03) $\times$ 10$^{-11}$ photons cm$^{-2}$ s$^{-1}$ with the TS value of 33.02 (5.7$\sigma$). To obtain the SED of SrcX, the data in the 30 GeV-1 TeV energy range were binned into five logarithmically equal intervals. And for each energy bin, we performed the same likelihood fitting analysis. In addition, we have calculated the upper limit with a confidence level of 95 per cent when the TS value of the energy bin is less than 4.0. As shown in Fig~\ref{Fig2}, we find that the SrcX SED has a hard GeV $\gamma$-ray spectrum, data listed in Table~\ref{tab2}.

\begin{figure*}
\centering
\begin{minipage}[t]{0.48\textwidth}
\centering
\includegraphics[height=6.3cm,width=8.0cm]{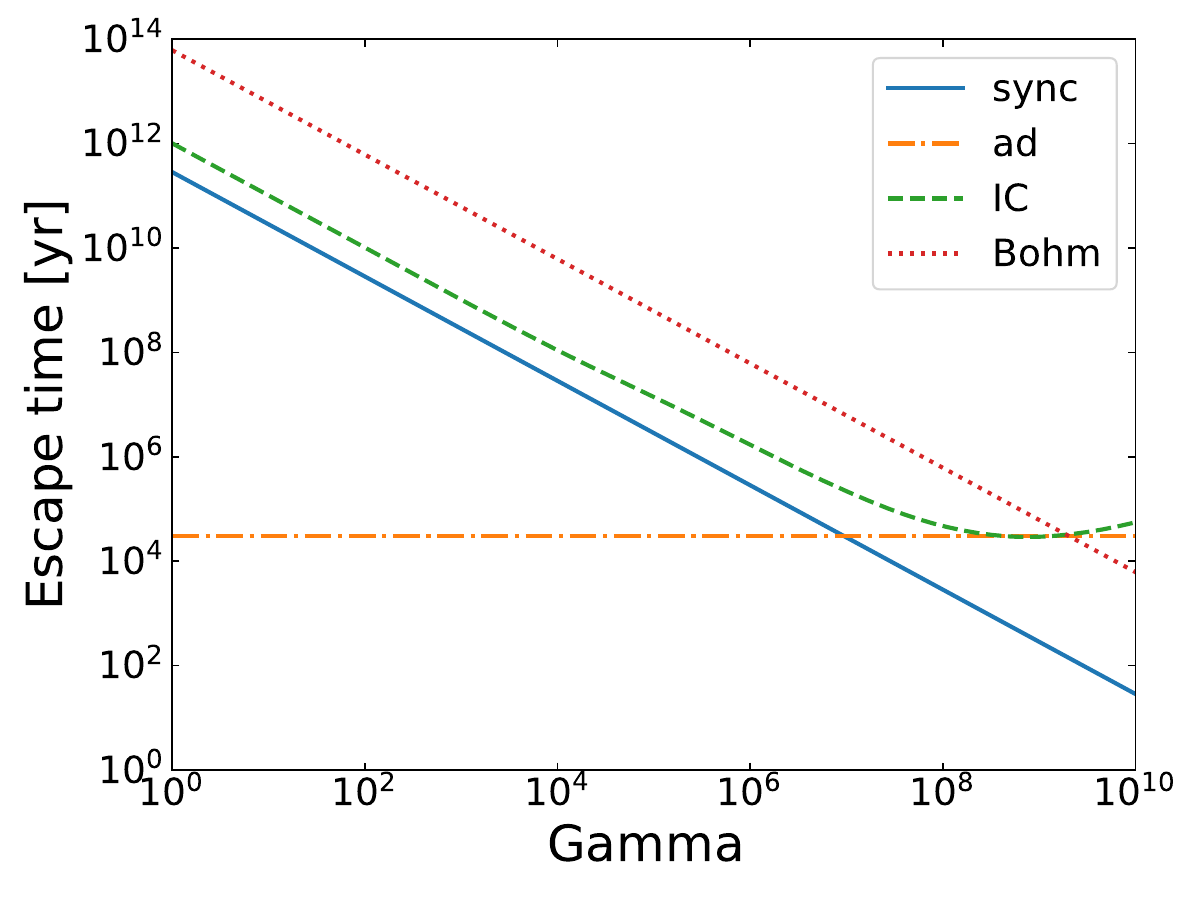}
\end{minipage}
\begin{minipage}[t]{0.48\textwidth}
\centering
\includegraphics[height=6.3cm,width=8.2cm]{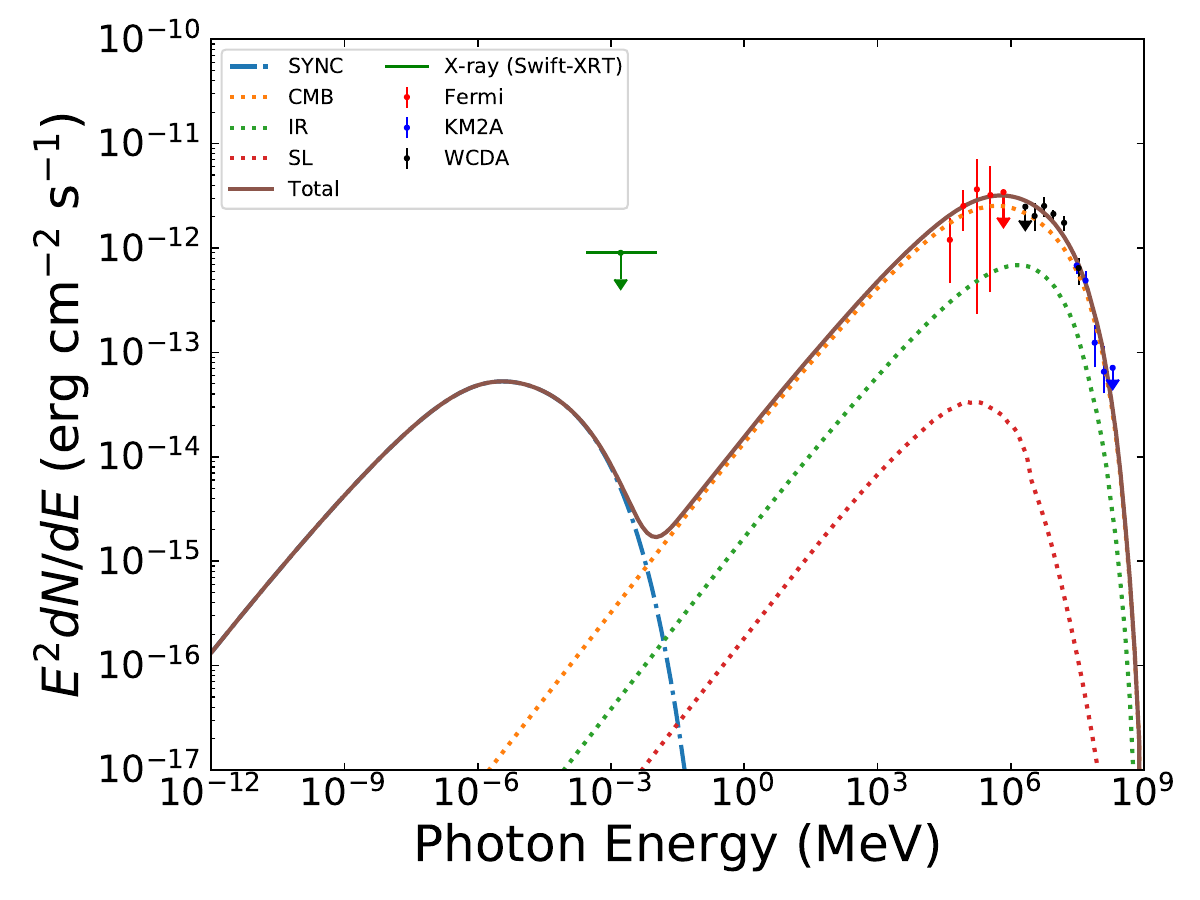}
\end{minipage}
\caption{Left panel: the cooling time scales for the synchrotron radiation (solid line), the adiabatic loss (dash-dotted line), the inverse Compton scattering (dashed line), and the escape time scale for the Bohm diffusion (dotted line) with $\varepsilon = 0.10$ at 30 kyr. Right panel: a comparison of the calculated multiband SED with the observed data for 1LHAASO J0249+6022 (model A). The dash-dotted, dotted, and solid lines are for the synchrotron, inverse Compton scattering off CMB, IR, and starlight, and the total emission, respectively. The red data points represent the results of this work, while the black and blue data points are taken from \cite{2024arXiv241004425C}, and the X-ray upper limit is taken from \cite{2011ApJ...733...82M} with the Swift-XRT data.
}
\label{Fig3}
\end{figure*}

\begin{table*}
   \centering
   \caption{Fitting results for different model parameters}
   \label{tab3}
   \renewcommand\arraystretch{1.0}
  \setlength{\tabcolsep}{2.8mm}
   \begin{tabular}{cccccccccccccc} 
      \hline \hline
Model & A & B & C & D & E& F  & G  & H & I & J & K & L &M   \\\hline
  $\alpha$& 1.9 & 1.7 & 1.8 & 2.0 & 2.1& 1.9  & 1.9  & 1.9 & 1.9 & 1.9 & 1.9 & 1.9 &1.9  \\
$\varepsilon$ & 0.10 & 0.10 & 0.10 & 0.10 & 0.10& 0.05  & 0.15  & 0.20 & 0.25 & 0.10 & 0.10 & 0.10 &0.10 \\
$\eta$ 	& 0.07 & 0.07 & 0.07 & 0.07 & 0.07& 0.07  & 0.07 & 0.07 & 0.07 & 0.02 & 0.12 & 0.17 &0.22 \\
B($\mu$G) 	& 9.26 & 9.26 & 9.26 & 9.26 & 9.26& 9.26  & 9.26  & 9.26 & 9.26 & 4.95 & 12.13 & 14.44 &16.42 \\
$\mathrm{E_{max}}$(PeV)	& 0.47 & 0.55 & 0.49 & 0.43 & 0.34& 0.24  & 0.69  & 0.88 & 0.99 & 0.30 & 0.55 & 0.62 &0.70 \\
$\chi^2$(dof=13)	& 0.82 & 33.68 & 8.42 & 4.45 & 9.86& 3.73  & 3.78  & 9.34 & 15.75 & 7.98 & 2.11 & 3.84 &5.35 \\\hline
\multicolumn{4}{l}{}
   \end{tabular}
\end{table*}

\section{Discussion}
\label{results}
The Fermi-LAT data analysis above shows that the excess of the GeV $\gamma$-ray emission around PSR J0248+6021 is spatially consistent with 1LHAASO J0249+6022, and the GeV spectrum of SrcX is also smoothly connects to the TeV spectrum of 1LHAASO J0249+6022 (see Fig~\ref{Fig3}). In the study of extended GeV emission, cases of sources of unknown nature with hard GeV spectral indices have been reported, e.g. HESS J1640-465 \citep{2018ApJ...867...55X}, VER J2227+608 \citep{2019ApJ...885..162X}, HESS J1837-069 \citep{2020MNRAS.494.3405S}, G17.8+16.7 \citep{2022MNRAS.510.2920A}, G118.4+37.0 \citep{2023MNRAS.518.4132A}, and HESS J1813-178 \citep{2024A&A...686A.149H}. After analysis, most sources are considered more likely to be associated with PWNe. The origin of the $\gamma$-ray emission from 1LHAASO J0249+6022 is currently unknown, but efforts have been made by \cite{2024arXiv241004425C}. Assuming that the distribution of the parent particles follows an exponential cutoff power law or a broken power law, \cite{2024arXiv241004425C} found that both the inverse Compton scattering process and the proton-proton inelastic collision fit the observed data well with reasonable energy budgets. However, high energy proton spectral indices of -1 (or even harder) are required in hadronic scenarios, which are difficult to achieve in the shock acceleration of SNR. In addition, hard spectra can be produced by transport effects if the $\gamma$-ray emission comes from a molecular cloud illuminated by propagating cosmic rays, but the molecular gas only partially overlaps the region of $\gamma$-ray emission and its centre of density is shifted away from the best-fit position of the $\gamma$-ray emission compared to the position of the pulsar based on the CO observations of \cite{2001ApJ...547..792D}. It is worth noting that the high energy electron spectral index of -1 or even harder is natural for the electrons injected by PWNe. As a result, \cite{2024arXiv241004425C} concluded that it is more likely that the extended $\gamma$-rays are associated with the TeV pulsar halo or the PWN. In the pulsar halo scenario, the diffusion coefficient is estimated to be $2 \times 10^{28}(d/2.0 \mathrm{kpc})^{2} \mathrm{cm}^{2}/\mathrm{s}$ for electrons/positrons at $\sim$160 TeV, which is close to the value of Monogem, but much larger than that of Geminga and PSR J0622+3749 \citep{2024arXiv241004425C}. Moreover, the diffusion coefficient is smaller than the Galactic one \citep{2017Sci...358..911A,2017PhRvD..96j3016L}. Here we focus on PWN scenarios for $\gamma$-ray radiation.

A leptonic time-dependent model of PWNe is used in this work. In the model, the time-dependent lepton population is balanced by injection, energy losses, and escape. The diffusion equation can be described as \citep[e.g.][]{2010A&A...515A..20F,2012MNRAS.427..415M}
\begin{equation}
\frac{\partial N(\gamma,t)}{\partial t}=-\frac{\partial}{\partial\gamma}[\dot{\gamma}(\gamma,t)N(\gamma,t)]-\frac{N(\gamma,t)}{\tau(\gamma,t)}+Q(\gamma,t) ,
\label{eq1}
\end{equation}
where the left-hand side is the variation of the lepton distribution in time. The function $\dot{\gamma}(\gamma,t)$ is the summation of the energy losses due to synchrotron, (Klein-Nishina) inverse Compton, self-synchrotron Compton, and adiabatic expansion. $Q(\gamma,t)$ is the injection of particles per unit energy per unit time, and $\tau(\gamma,t)$ is the escape time due to Bohm diffusion \citep{2008ApJ...676.1210Z}.

\begin{figure*}
\centering
\begin{minipage}[t]{0.45\textwidth}
\centering
\includegraphics[height=6.0cm,width=8.0cm]{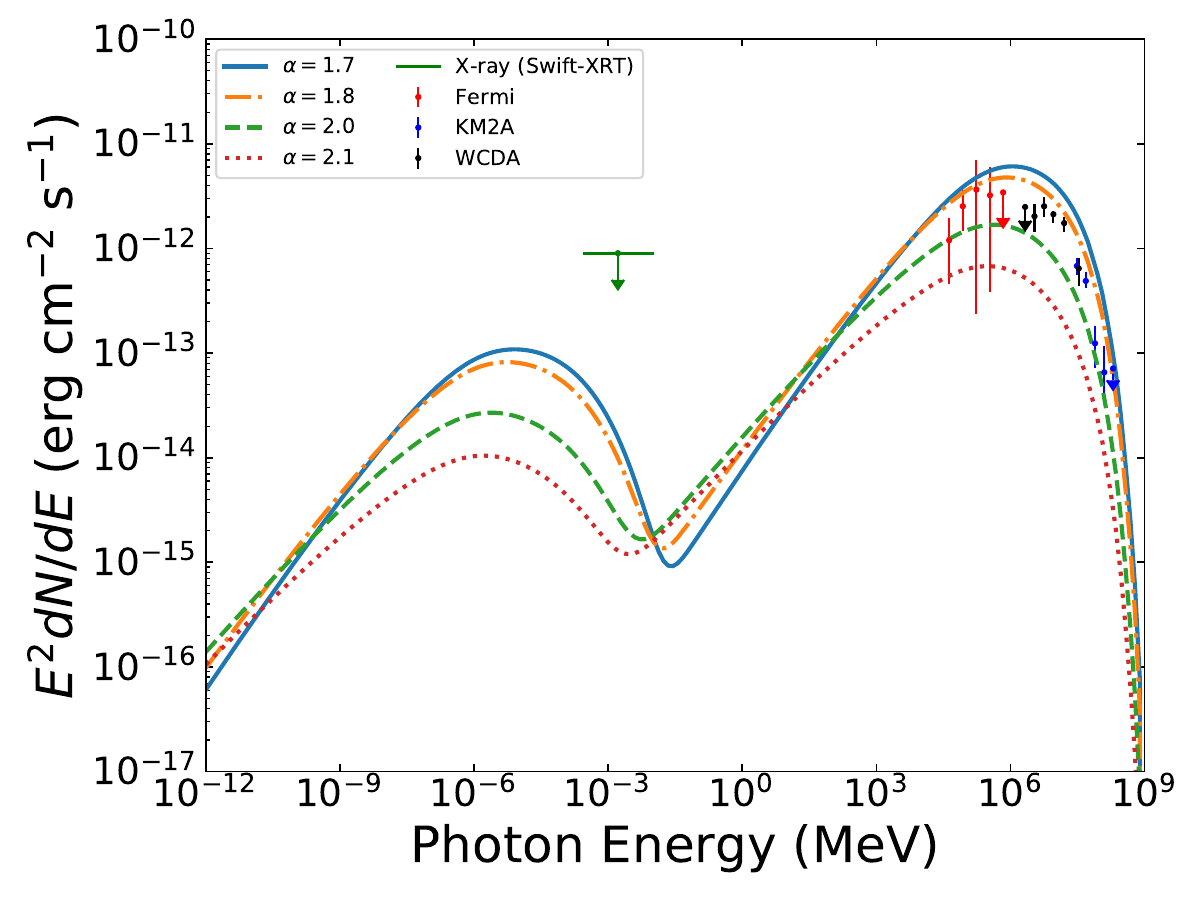}
\end{minipage}
\begin{minipage}[t]{0.38\textwidth}
\centering
\includegraphics[height=6.0cm,width=8.0cm]{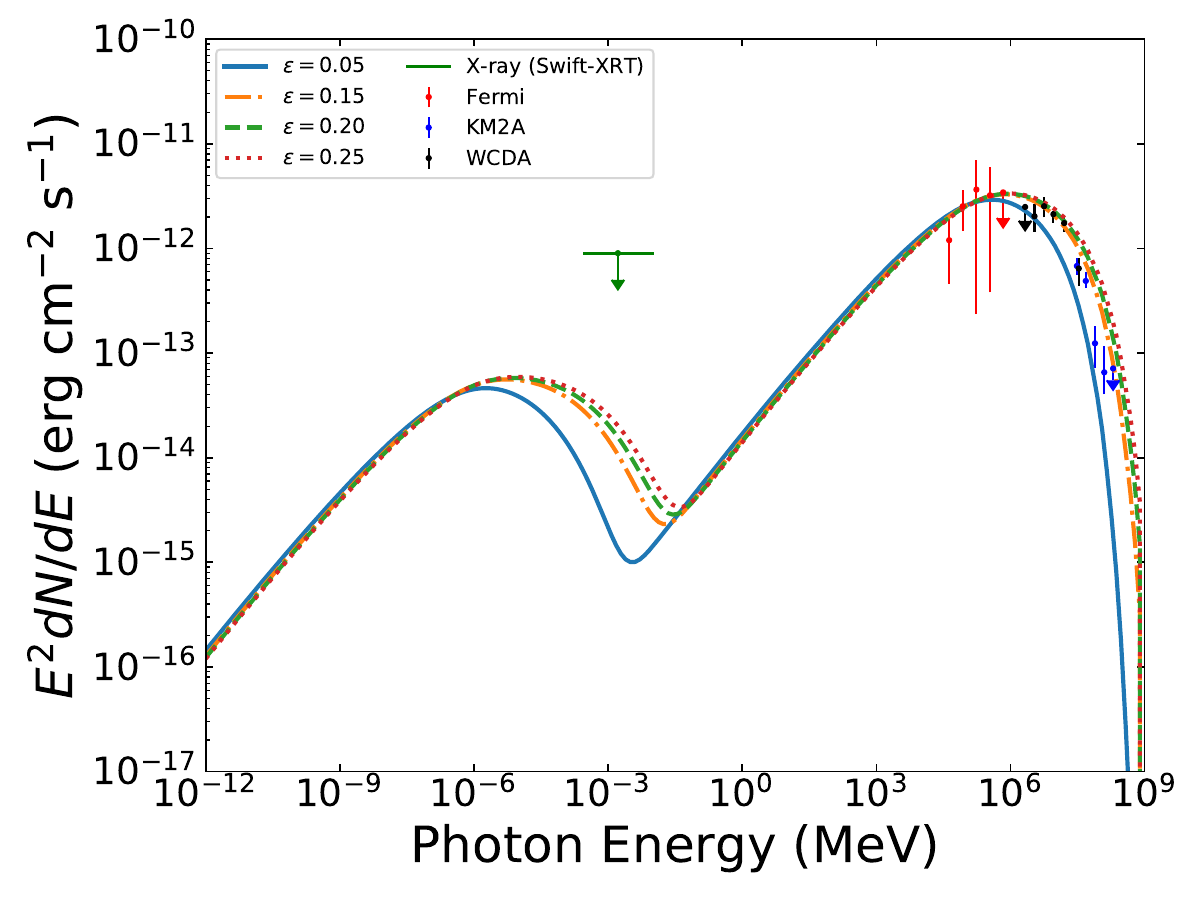}
\end{minipage}
\begin{minipage}[t]{0.40\textwidth}
\centering
\includegraphics[height=6.0cm,width=8.0cm]{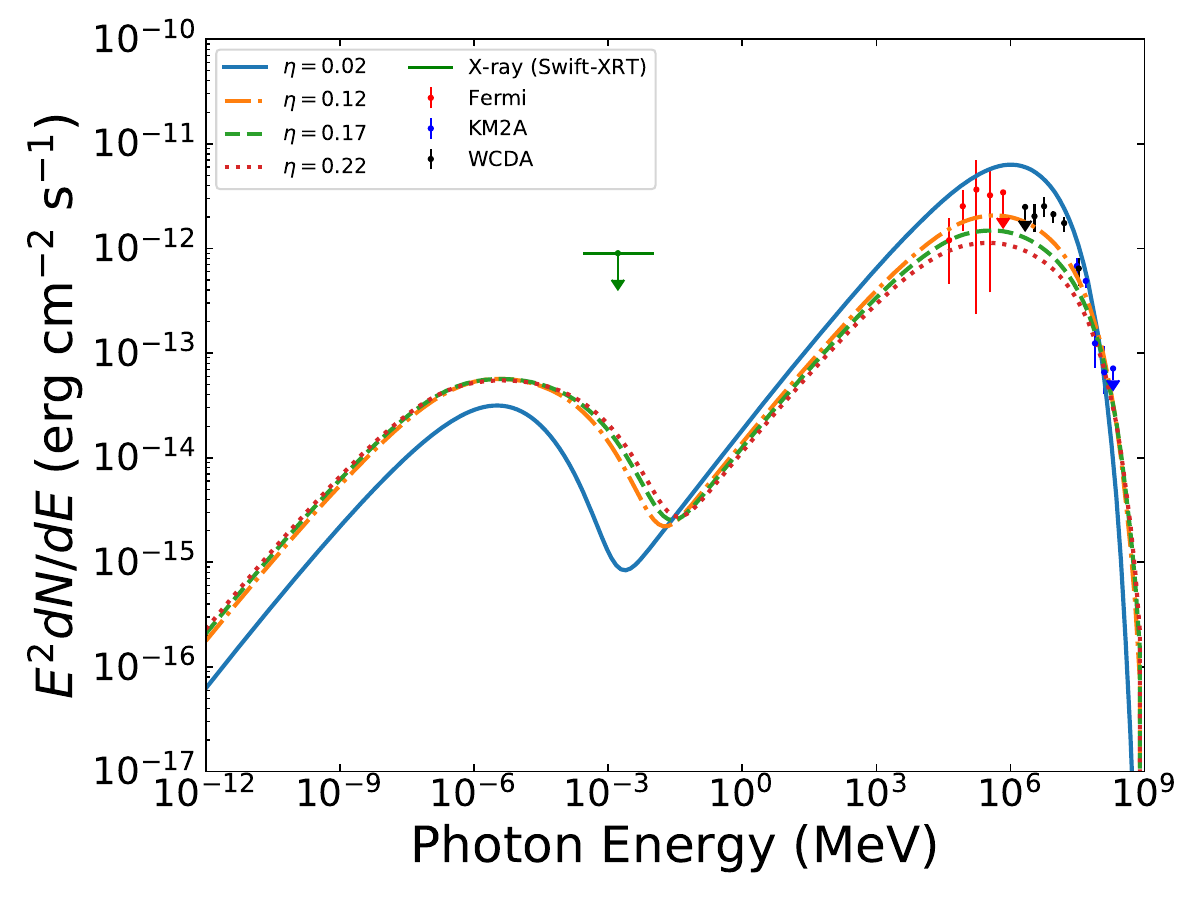}
\end{minipage}
\caption{Only one model parameter is changed to fit the observed fluxes in multiple bands, while the other parameters are the same as in model A. Upper left: the resulting SEDs with different values of $\alpha$ (model B-E). Upper right: the resulting SEDs with different values of $\varepsilon$ (model F-I). Bottom: the resulting SEDs with different values of $\eta$ (model J-M).
}
\label{Fig4}
\end{figure*}

Any form of particle injection is possible in the numerical implementation of \cite{2012MNRAS.427..415M}. We assume that leptons are continuously injected into the PWN with a power-law distribution \citep{1984ApJ...283..694K,2020MNRAS.498.4901F}
\begin{equation}
Q(\gamma,t)=Q_0(t)\gamma^{-\alpha},
\label{eq2}
\end{equation}
The normalization constant $Q_0(t)$ is written in the form \citep{2014JHEAp...1...31T}
\begin{equation}
(1-\eta)L(t)=\int\gamma m_\text{e}c^2Q(\gamma,t)\text{d}\gamma,
\label{eq3}
\end{equation}
The injection luminosity $L(t)$ is determined by
\begin{equation}
L(t)=L_0\left(1+\frac{t}{\tau_0}\right)^{-\frac{n+1}{n-1}},
\label{eq4}
\end{equation}
where $L_0$ and $n$ are the initial luminosity and the breaking index (usually taken as 3), respectively. The initial spin-down time-scale of the pulsar is
\begin{equation}
\tau_0=\frac{P_0}{(n-1)\dot{P}_0}=\frac{2\tau_\mathrm{c}}{n-1}-t_\mathrm{age},
\label{eq5}
\end{equation}
where $P_0$ and $\dot{P}_0$ are the initial period and its first derivative and $\tau_\mathrm{c}=P/2\dot{P}$ is the characteristic age of the pulsar. From the period ($P$) and first derivative ($\dot{P}$) of the pulsar, we can obtain the spin-down luminosity
\begin{equation}
L(t)=4\pi^2I \frac{P}{\dot{P}^3},
\label{eq6}
\end{equation}
where $I$ is the pulsar moment of inertia, which we assume $I\sim10^{45} \mathrm{g~cm}^{2}$. The spin-down luminosity is transferred into particles (mostly electrons and positrons) and magnetic field. Assuming the magnetic energy fraction is $\eta$, the average magnetic field in the nebula is calculated with \citep{2010ApJ...715.1248T,2012MNRAS.427..415M}
\begin{equation}
B(t)=\sqrt{\frac{3(n-1)\eta L_0\tau_0}{R_{\text{PWN}}^3(t)}\left[1-\left(1+\frac{t}{\tau_0}\right)^{-\frac{2}{n-1}}\right]},
\label{eq6}
\end{equation}
The expansion behaviour of the PWN within the SNR is determined by the age of the system $t$, the initial spin-down timescale $\tau_0$, and the reverse-shock interaction time $t_{rs}$. The radius of the nebula is \citep{2018A&A...612A...2H}
\begin{equation}
R(t)\quad\propto\quad\begin{cases}t^{6/5}&\text{for} \ t\leqslant\tau_0,\\t&\text{for} \ \tau_0<t\leqslant t_\text{rs},\\t^{3/10}&\text{for} \ t>t_\text{rs}.\end{cases}
\label{eq7}
\end{equation}
To ensure particle confinement, we impose that the Larmor radius $R_{\mathrm{L}}$ must be less than a fraction of the radius of the termination shock $R_{\mathrm{s}}$, which leads to
\begin{equation}
\gamma_{\max}(t)=\frac{\varepsilon e\kappa}{m_{\mathrm{e}}c^2}\left(\frac{\eta L(t)}{c}\right)^{1/2}.
\label{eq10}
\end{equation}
where $\gamma_{\max}$, $e$, and $m_{\mathrm{e}}$ are the maximum Lorentz factor, the electron mass, and charge, respectively. $\varepsilon$ represents the fractional size of the shock radius, which must be less than 1 due to $R_{\mathrm{L}}=\varepsilon R_{\mathrm{s}}$. $\kappa$ denotes the magnetic compression ratio, which equals about 3 for strong shocks \citep{2012A&A...539A..24H}.

We assume that the 1LHAASO J0249+6022 is powered by the pulsar PSR J0248+6021 with $P = 0.22 s$, $\dot{P} = 5.51\times10^{-14}\text{s s}^{-1}$,  $\tau_\mathrm{c}=63.2$ kyr, and a distance of d = 2.0 kpc \citep{2011ApJ...733...82M,2011A&A...525A..94T}. The age of PSR J0248+6021 is unknown, we assume it to be 30 kyr. With this age, the initial spin-down time-scale $\tau_\mathrm{0} = 3.24 \times 10^4$ yr was evaluated with equation~\ref{eq5}, and the initial luminosity $L_0 = 1.96 \times 10^{38} \mathrm{erg \ s^{-1}}$ was evaluated with equation~\ref{eq4}. For the observed multiband data, the X-ray upper limits is taken from \cite{2011ApJ...733...82M}, Fermi-LAT data are taken from the analysed results of this work, and TeV data are taken from \cite{2024arXiv241004425C}. With regard to the process of data extraction, photons within a circle of radius $\mathrm{20''}$ centered on the pulsar were selected for the X-ray data, photons within a circle of radius $0.54^{\circ}$ centered on the SrcX best-fit position were selected for the Fermi data, photons within a circle of radius $0.69^{\circ}$ centered on the WCDA position were selected for the WCDA data, and photons within a circle of radius $0.37^{\circ}$ centered on the KM2A position were selected for the KM2A data.

The SEDs of non-thermal photons are calculated through synchrotron radiation and inverse Compton scattering off the background soft photons, which consist of the synchrotron emission, cosmic-microwave background (CMB), infrared (IR) photons, and starlight (SL). We used the radiative transfer model of \cite{2017MNRAS.470.2539P} to estimate the energy density of the field. The energy densities U$_\mathrm{{CMB}}$ = 0.25 eV cm$^{-3}$, U$_\mathrm{{IR}}$ = 0.12 eV cm$^{-3}$, and U$_\mathrm{{SL}}$ = 0.23 eV cm$^{-3}$, and temperatures T$_\mathrm{{CMB}}$ = 2.7 K, T$_\mathrm{{IR}}$ = 35.0 K, and T$_\mathrm{{SL}}$ = 6500.0 K for the CMB, IR, and SL photons are used in the calculation.

According to previous research, it was found that the energies contained in the particles and the magnetic field do not satisfy the equipartition \citep{2014JHEAp...1...31T,2018A&A...609A.110Z,2024ApJ...967..127Z}. A magnetic energy ratio $\eta$ of less than 0.1 can effectively reproduce multiband observational data in PWNe model calculations. With $\eta = 0.07$, the magnetic field strength in the nebula is calculated to be 9.3 $\mu$G at a time of 30 kyr. To research the effects of the various physical processes, the different cooling timescales and escape timescales were calculated. As shown in Fig.~\ref{Fig3}, the left panel represents the cooling times for the adiabatic loss, the synchrotron radiation, the inverse Compton scattering, and the escape time due to the Bohm diffusion for the particles with different Lorentz factor. The effects of the cooling mechanisms are dominated by adiabatic loss for $\gamma < 10^7$, the synchrotron loss dominate over the inverse Compton scattering losses and Bohm diffusion for $\gamma > 10^7$. Assuming that the particles injected into the nebula have a power-law distribution with $\alpha = 1.9$, $\varepsilon = 0.1$, and $\eta = 0.07$ (model A), the resulting SED of the 1LHAASO J0249+6022 at $\mathrm{T_{age}}$ = 30 kyr is shown in the right panel of Fig.~\ref{Fig3}. It can be observed that the detected fluxes of the multi-band non-thermal emission can be well reproduced with the model. As shown in the right panel of Fig.~\ref{Fig3}, the inverse Compton scattering off the CMB is the strongest contributor to the high-energy spectra, followed by inverse Compton with the IR and the SL. In addition, our model indicates a nebula radius of 10.0 pc, which is consistent with the physical dimensions derived using the Gaussian template assumption of \cite{2024arXiv241004425C}. Finally, the maximum energy of the injected electrons and positrons is calculated to be 0.47 PeV.

In SED modelling, different combinations of parameters result in different SED outcomes. We tested the effect of variations in the three parameters ($\alpha$, $\varepsilon$, and $\eta$) of the model on SED. As shown in Fig.~\ref{Fig4} and Table~\ref{tab3}, the magnetic field B increases solely in conjunction with parameter $\eta$ (while maintaining all other parameters constant), and the maximum energy of the particles $\mathrm{E_{max}}$ exhibits an increase in conjunction with parameter $\eta$ or parameter $\varepsilon$, while concurrently decreasing in conjunction with parameter $\alpha$. In addition, we find that model A has the smallest reduced $\chi^2$ value.
\section{Summary}
\label{sumdis}
In this work, we analyzed the GeV $\gamma$-ray emission in the field of 1LHAASO J0249+6022, using $\sim$16 yr of Fermi-LAT data, and found an extended $\gamma$-ray source SrcX, which can be described by a two-dimensional Gaussian template. The GeV $\gamma$-ray spectrum of SrcX follows a hard power-law model with an index of 1.54 $\pm$ 0.17 in the 30 GeV-1 TeV energy range, which can connect with the TeV SED of 1LHAASO J0249+6022 smoothly. The spatial and spectral associations indicate that SrcX could be the GeV counterpart of 1LHAASO J0249+6022. We then investigated whether the $\gamma$-ray emission from 1LHAASO J0249+6022 could be the result of the PWN scenario with a one-zone time-dependent model. Assuming the electrons/positrons injected into the nebula has the power-law distribution, the resulting SED of the model is consistent with the fluxes detected by Fermi-LAT and LHAASO. Our study supports that the nebula powered by the energetic pulsar PSR J0248+6021 is capable of producing the $\gamma$-ray flux of 1LHAASO J0249+6022, and that electrons/positrons injected into the nebula can be accelerated up to $\sim$0.47 PeV. Future high-resolution and multiwavelength observations are expected to yield further insights into the origin of the high-energy particles observed in this source, as well as the properties of the interstellar medium in this region.

\section*{Acknowledgements}
{We thank the anonymous referee for comments that helped improve this work. This work was partially supported by the National Natural Science Foundation of China (12233006, 12063004, 12393852, 12103046). This research is supported by the Yunnan Provincial Government (YNWR-QNBJ-2018-049) and Yunnan Fundamental Research Projects (grant No. 202201BF070001-020, 202101AU070036).
}


\begin{thebibliography}{}
\bibitem[Abdo et al.(2010)]{2010ApJS..187..460A} Abdo, A.~A., Ackermann, M., Ajello, M., et al.\ 2010, \apjs, 187, 460. doi:10.1088/0067-0049/187/2/460
\bibitem[Abdollahi et al.(2020)]{2020ApJS..247...33A} Abdollahi, S., Acero, F., Ackermann, M., et al.\ 2020, \apjs, 247, 33. doi:10.3847/1538-4365/ab6bcb
\bibitem[Abdollahi et al.(2022)]{2022ApJS..260...53A} Abdollahi, S., Acero, F., Baldini, L., et al.\ 2022, \apjs, 260, 53. doi:10.3847/1538-4365/ac6751
\bibitem[Abe et al.(2023)]{2023A&A...673A..75A} Abe, S., Aguasca-Cabot, A., Agudo, I., et al.\ 2023, \aap, 673, A75. doi:10.1051/0004-6361/202245086
\bibitem[Abeysekara et al.(2017)]{2017Sci...358..911A} Abeysekara, A.~U., Albert, A., Alfaro, R., et al.\ 2017, Science, 358, 911. doi:10.1126/science.aan4880
\bibitem[Akaike(1974)]{1974ITAC...19..716A} Akaike, H.\ 1974, IEEE Transactions on Automatic Control, 19, 716
\bibitem[Albert et al.(2020)]{2020ApJ...905...76A} Albert, A., Alfaro, R., Alvarez, C., et al.\ 2020, \apj, 905, 76. doi:10.3847/1538-4357/abc2d8
\bibitem[Araya et al.(2022)]{2022MNRAS.510.2920A} Araya, M., Hurley-Walker, N., \& Quir{\'o}s-Araya, S.\ 2022, \mnras, 510, 2920. doi:10.1093/mnras/stab3550
\bibitem[Araya(2023)]{2023MNRAS.518.4132A} Araya, M.\ 2023, \mnras, 518, 4132. doi:10.1093/mnras/stac3337
\bibitem[Araya \& {\'A}lvarez-Quesada(2024)]{2024MNRAS.527.8006A} Araya, M. \& {\'A}lvarez-Quesada, J.~A.\ 2024, \mnras, 527, 8006. doi:10.1093/mnras/stad3739
\bibitem[Atwood et al.(2009)]{2009ApJ...697.1071A} Atwood, W.~B., Abdo, A.~A., Ackermann, M., et al.\ 2009, \apj, 697, 1071. doi:10.1088/0004-637X/697/2/1071
\bibitem[Ballet et al.(2023)]{2023arXiv230712546B} Ballet, J., Bruel, P., Burnett, T.~H., et al.\ 2023, arXiv:2307.12546. doi:10.48550/arXiv.2307.12546
\bibitem[Cao et al.(2021)]{2021Natur.594...33C} Cao, Z., Aharonian, F.~A., An, Q., et al.\ 2021, \nat, 594, 33. doi:10.1038/s41586-021-03498-z
\bibitem[Cao et al.(2024a)]{2024ApJS..271...25C} Cao, Z., Aharonian, F., An, Q., et al.\ 2024a, \apjs, 271, 25. doi:10.3847/1538-4365/acfd29
\bibitem[Cao et al.(2024b)]{2024arXiv241004425C} Cao, Z., Aharonian, F., An, Q., et al.\ 2024b, arXiv:2410.04425. doi:10.48550/arXiv.2410.04425
\bibitem[Dame et al.(2001)]{2001ApJ...547..792D} Dame, T.~M., Hartmann, D., \& Thaddeus, P.\ 2001, \apj, 547, 792. doi:10.1086/318388
\bibitem[Eagle et al.(2020)]{2020ApJ...904..123E} Eagle, J., Marchesi, S., Castro, D., et al.\ 2020, \apj, 904, 123. doi:10.3847/1538-4357/abbe08
\bibitem[Fang \& Zhang(2010)]{2010A&A...515A..20F} Fang, J. \& Zhang, L.\ 2010, \aap, 515, A20. doi:10.1051/0004-6361/200913615
\bibitem[Fang et al.(2020)]{2020MNRAS.498.4901F} Fang, J., Wen, L., Yu, H., et al.\ 2020, \mnras, 498, 4901. doi:10.1093/mnras/staa2703
\bibitem[Foster et al.(1997)]{1997AAS...19111110F} Foster, R.~S., Ray, P.~S., Cadwell, B.~J., et al.\ 1997, \aas, 191, 111.10
\bibitem[Guo \& Xin(2024)]{2024ApJ...965...28G} Guo, X. \& Xin, Y.\ 2024, \apj, 965, 28. doi:10.3847/1538-4357/ad2ae1
\bibitem[Gong et al.(2023)]{2023MNRAS.526..193G} Gong, Y., Xiao, Y., Zhou, L., et al.\ 2023, \mnras, 526, 193. doi:10.1093/mnras/stad2649
\bibitem[Holler et al.(2012)]{2012A&A...539A..24H} Holler, M., Sch{\"o}ck, F.~M., Eger, P., et al.\ 2012, \aap, 539, A24. doi:10.1051/0004-6361/201118121
\bibitem[H.~E.~S.~S. Collaboration et al.(2018)]{2018A&A...612A...1H} H.~E.~S.~S. Collaboration, Abdalla, H., Abramowski, A., et al.\ 2018, \aap, 612, A1. doi:10.1051/0004-6361/201732098
\bibitem[H.~E.~S.~S. Collaboration et al.(2018)]{2018A&A...612A...2H} H.~E.~S.~S. Collaboration, Abdalla, H., Abramowski, A., et al.\ 2018, \aap, 612, A2. doi:10.1051/0004-6361/201629377
\bibitem[H.~E.~S.~S. Collaboration et al.(2024)]{2024A&A...686A.149H} H.~E.~S.~S. Collaboration, Aharonian, F., Ait Benkhali, F., et al.\ 2024, \aap, 686, A149. doi:10.1051/0004-6361/202348374
\bibitem[Kennel \& Coroniti(1984)]{1984ApJ...283..694K} Kennel, C.~F. \& Coroniti, F.~V.\ 1984, \apj, 283, 694. doi:10.1086/162356
\bibitem[Kharchenko et al.(2005)]{2005A&A...440..403K} Kharchenko, N.~V., Piskunov, A.~E., R{\"o}ser, S., et al.\ 2005, \aap, 440, 403. doi:10.1051/0004-6361:20052740
\bibitem[Lande et al.(2012)]{2012ApJ...756....5L} Lande, J., Ackermann, M., Allafort, A., et al.\ 2012, \apj, 756, 5. doi:10.1088/0004-637X/756/1/5
\bibitem[Li et al.(2021)]{2021ApJ...913L..33L} Li, J., Liu, R.-Y., de O{\~n}a Wilhelmi, E., et al.\ 2021, \apjl, 913, L33. doi:10.3847/2041-8213/abf925
\bibitem[Linden et al.(2017)]{2017PhRvD..96j3016L} Linden, T., Auchettl, K., Bramante, J., et al.\ 2017, \prd, 96, 103016. doi:10.1103/PhysRevD.96.103016
\bibitem[MAGIC Collaboration et al.(2020)]{2020A&A...643L..14M} MAGIC Collaboration, Acciari, V.~A., Ansoldi, S., et al.\ 2020, \aap, 643, L14. doi:10.1051/0004-6361/202039131
\bibitem[Marelli et al.(2011)]{2011ApJ...733...82M} Marelli, M., De Luca, A., \& Caraveo, P.~A.\ 2011, \apj, 733, 82. doi:10.1088/0004-637X/733/2/82
\bibitem[Mart{\'\i}n et al.(2012)]{2012MNRAS.427..415M} Mart{\'\i}n, J., Torres, D.~F., \& Rea, N.\ 2012, \mnras, 427, 415. doi:10.1111/j.1365-2966.2012.22014.x
\bibitem[Popescu et al.(2017)]{2017MNRAS.470.2539P} Popescu, C.~C., Yang, R., Tuffs, R.~J., et al.\ 2017, \mnras, 470, 2539. doi:10.1093/mnras/stx1282
\bibitem[Sun et al.(2020)]{2020MNRAS.494.3405S} Sun, X.-N., Yang, R.-Z., \& Wang, X.-Y.\ 2020, \mnras, 494, 3405. doi:10.1093/mnras/staa947
\bibitem[Tanaka \& Takahara(2010)]{2010ApJ...715.1248T} Tanaka, S.~J. \& Takahara, F.\ 2010, \apj, 715, 1248. doi:10.1088/0004-637X/715/2/1248
\bibitem[Theureau et al.(2011)]{2011A&A...525A..94T} Theureau, G., Parent, D., Cognard, I., et al.\ 2011, \aap, 525, A94. doi:10.1051/0004-6361/201015317
\bibitem[Torres et al.(2014)]{2014JHEAp...1...31T} Torres, D.~F., Cillis, A., Mart{\'\i}n, J., et al.\ 2014, Journal of High Energy Astrophysics, 1, 31. doi:10.1016/j.jheap.2014.02.001
\bibitem[Wang et al.(2020)]{2020A&A...644A..73W} Wang, P.~F., Han, J.~L., Han, L., et al.\ 2020, \aap, 644, A73. doi:10.1051/0004-6361/202038867
\bibitem[Xiang et al.(2021)]{2021ApJ...911...49X} Xiang, Y., Jiang, Z., \& Tang, Y.\ 2021, \apj, 911, 49. doi:10.3847/1538-4357/abeb19
\bibitem[Xiang et al.(2021)]{2021ApJ...912..117X} Xiang, Y., Xing, Y., \& Jiang, Z.\ 2021, \apj, 912, 117. doi:10.3847/1538-4357/abe624
\bibitem[Xiang \& Jiang(2021)]{2021ApJ...918...24X} Xiang, Y. \& Jiang, Z.\ 2021, \apj, 918, 24. doi:10.3847/1538-4357/ac0fe4
\bibitem[Xiao et al.(2024)]{2024ApJ...972...84X} Xiao, Y., Wu, K., \& Fang, J.\ 2024, \apj, 972, 84. doi:10.3847/1538-4357/ad6563
\bibitem[Xin et al.(2018)]{2018ApJ...867...55X} Xin, Y.-L., Liao, N.-H., Guo, X.-L., et al.\ 2018, \apj, 867, 55. doi:10.3847/1538-4357/aae313
\bibitem[Xin et al.(2019)]{2019ApJ...885..162X} Xin, Y., Zeng, H., Liu, S., et al.\ 2019, \apj, 885, 162. doi:10.3847/1538-4357/ab48ee
\bibitem[Zheng \& Wang(2024)]{2024ApJ...968..117Z} Zheng, D. \& Wang, Z.\ 2024, \apj, 968, 117. doi:10.3847/1538-4357/ad496d
\bibitem[Zhang et al.(2008)]{2008ApJ...676.1210Z} Zhang, L., Chen, S.~B., \& Fang, J.\ 2008, \apj, 676, 1210. doi:10.1086/527466
\bibitem[Zhou et al.(2024)]{2024MNRAS.529.3593Z} Zhou, L., Wu, K., Gong, Y., et al.\ 2024, \mnras, 529, 3593. doi:10.1093/mnras/stae720
\bibitem[Zhu et al.(2018)]{2018A&A...609A.110Z} Zhu, B.-T., Zhang, L., \& Fang, J.\ 2018, \aap, 609, A110. doi:10.1051/0004-6361/201629108
\bibitem[Zhu et al.(2024)]{2024ApJ...967..127Z} Zhu, B.-T., Lu, F.-W., \& Zhang, L.\ 2024, \apj, 967, 127. doi:10.3847/1538-4357/ad445e
\end{thebibliography}
\end{document}